\documentclass[a4paper,12pt,onecolumn]{article}
\usepackage{ar} 
\usepackage{color}
\usepackage[pdftex]{graphicx}
\usepackage{epsfig}
\usepackage{bm}% bold math
\usepackage{dcolumn}% Align table columns on decimal point
\usepackage{amsmath, amssymb}
\usepackage{float}
\usepackage[hidelinks]{hyperref}

\usepackage[numbers,super,sort&compress]{natbib}

\usepackage{afterpage}

 \title{A unified slip boundary condition for flow over a surface}
 \date{\vspace{-5ex}}
 \author{
  Joseph J. Thalakkottor\thanks{Department of Mechanical and Aerospace Engineering, University of Florida, Gainesville, FL 32611.}
  \ and Kamran Mohseni$^*$\thanks{Department of Electrical and Computer Engineering, University of Florida, Gainesville, FL 32611.}
  }

\begin{document}

\maketitle
\begin{abstract}
 Interface between two phases of matter are ubiquitous in nature and technology. Determining the correct velocity condition at an interface is essential for understanding and designing of flows over a surface. We demonstrate that both the widely used no-slip and the Navier and Maxwell slip boundary conditions do not capture the complete physics associated with complex problems, such as spreading of liquids or corner flows. Hence, we present a unified boundary condition that is applicable to a wide-range of flow problems.
\end{abstract}

\section{Introduction}

\noindent { The interface between two phases of matter is often accompanied by rapid changes in scales, multi-physics, geometrical complexities and intriguing chemical phenomena, making it an ideal benchmark to expand our knowledge beyond the confines of the bulk material. Determining the correct matching boundary conditions is essential for accurate predictions, as they govern the transfer of mass, momentum and energy across such interfaces. Among these, the boundary condition governing the transfer of tangential momentum across a fluid-solid interface is a topic that is still being debated, despite over a century of scientific work \cite{GoldsteinS:38a,GoldsteinS:69a,VinogradovaOI:99a}. Existing models  \cite{NavierCLMH:1823a,MaxwellJC:90a,TroianSM:97a,QianT:06a} are unable to consistently explain and capture the complete physics associated with more complex problems such as a moving contact line (spreading of liquids) \cite{HuhC:71a,Hocking:82a,Dussan:79a,ThompsonP:89a} and corner flows \cite{MoffattHK:64a,KoplikJ:95a}.  
\textcolor{black}{Here we develop a unified boundary condition that is applicable for a wide range of fluid flow problems and includes the no-slip, Navier/Maxwell \cite{NavierCLMH:1823a,MaxwellJC:90a}, and Thompson \& Troian \cite{TroianSM:97a} velocity boundary conditions as limiting cases. The unified boundary condition, which is validated with molecular dynamics simulations, consists of two parts:}
First is the generalized velocity boundary condition, which accounts for the variation of flow velocity not only in the wall normal direction, as is the case for the Navier/Maxwell models, but also in the wall tangent direction. From this follows the second part where slip length is shown to be not just a constant, as suggested by Navier/Maxwell, nor a non-linear function of just the shear rate, as suggested by Thompson \& Troian, but rather a non-linear function of the principal strain rate. This universal relation for slip length along with the general velocity slip boundary condition provides a unified boundary condition to model a wide range of viscous flows over a solid surface.}

%%%%%%%%%%%%%%%%%%%%%%%%%%%%%%%%%%%%%%%%%%%%%%%%%%%%%%%%%%%%%%%%%%%%%%%%%%%%%%%%%%%%%%%%%%%%%%%%%%%%%%%%%%%%%%%%
%%%%%%%%%%%%%%%%%%%%%		Introduction / Setup / T&T validation 	%%%%%%%%%%%%%%%%%%%%%%%%%%%%%%%%%%%%%%%%
%%%%%%%%%%%%%%%%%%%%%%%%%%%%%%%%%%%%%%%%%%%%%%%%%%%%%%%%%%%%%%%%%%%%%%%%%%%%%%%%%%%%%%%%%%%%%%%%%%%%%%%%%%%%%%%%
\section{Verifying Thompson and Troian's model for moving contact line and corner flow problems}
\begin{figure}[h!]
\centering
\begin{minipage}{0.45\linewidth}\begin{center}
 \includegraphics[width=0.9\linewidth]{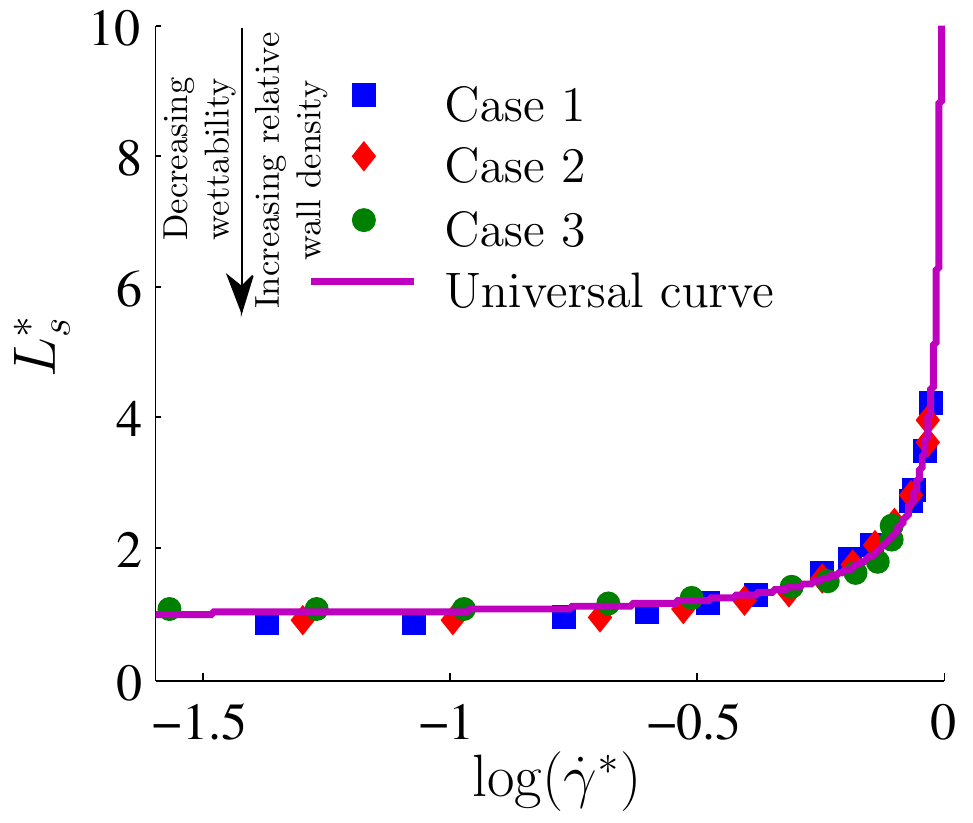}
\end{center}\end{minipage}
\begin{minipage}{0.45\linewidth}\begin{center}
 \includegraphics[width=0.9\linewidth]{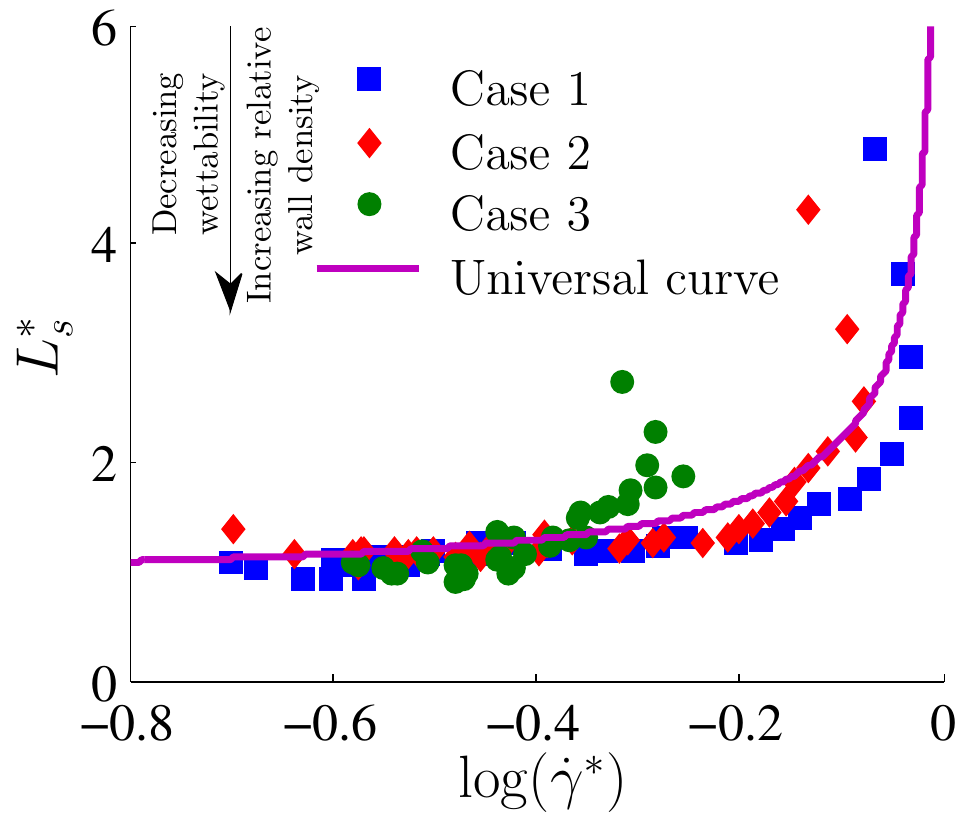}
\end{center}\end{minipage}\\
\begin{minipage}{0.49\linewidth}\begin{center} (a) unidirectional Couette flow \end{center}\end{minipage}
\begin{minipage}{0.49\linewidth}\begin{center} (b) moving contact line \end{center}\end{minipage}

\caption{{\bf Breakdown of Thompson \& Troian's slip model for a moving contact line problem.} Scaled slip length versus shear rate is plotted for (a) a single phase unidirectional Couette flow problem and (b) a two-phase, moving contact line problem. For both plots the data correspond to different fluid-wall properties, where hydrophobicity increases from Case 1 to 4. As predicted by Thompson \& Troian, by scaling slip length with $L_s^o$ and shear rate with $\dot{\gamma_c}$, the data for a single phase Couette flow collapses to a single curve. This is defined by the curve $L_s^*=(1-\dot{\gamma}^*)^{-0.5}$ where $L_s^*=L_s/L_s^o$, $\dot{\gamma}^*=\dot{\gamma}/\dot{\gamma_c}$ and $\dot{\gamma}=du/dy$ is the shear rate. However, the same scaling does not result in the collapse of data for moving contact line as seen in (b).}
\label{fig:ThndTr}
\end{figure}

The no-slip boundary condition is known to be valid for many continuum scale problems. However, in some cases, such as spreading of fluid on a solid surface \cite{HuhC:71a,Hocking:82a,Dussan:79a,ThompsonP:89a}, corner flow \cite{MoffattHK:64a,KoplikJ:95a} and extrusion of polymer melts from capillary tubes \cite{RichardsonS:73a}, assuming no-slip at the boundary leads to velocity and stress singularities, and the breakdown of the no-slip boundary condition. While steady flow boundary conditions for simple regular interfaces are fairly well understood, there is still a significant void in our understanding of the behaviour  near the intersection of multiple interfaces, such as a moving contact line (MCL) or a corner point. Here, the limiting factor is that the breakdown of the no-slip boundary condition at these intersections occurs at molecular scales. One of the proposed methods to alleviate these singularities is to assume fluid slip at these intersections. The two most common slip models are those presented more than a century ago, by Navier \cite{NavierCLMH:1823a} and Maxwell \cite{MaxwellJC:90a}. \textcolor{black}{However, Navier's and Maxwell's assumption of constant slip length for a given fluid-wall interface contradicts the experimental results \cite{ThompsonP:89a}, which show perfect slip at the singular point and no-slip far away from it.} Thompson \& Troian \cite{TroianSM:97a} suggested that their model resolves this, as it naturally allows for varying degrees of slip on approaching regions of high shear stress and shear rate. By scaling slip length ($L_s$) with its asymptotic value ($L_s^o$) and shear rate ($\dot{\gamma}$) by its critical value ($\dot{\gamma}_c$), they showed that the data for a steady Couette flow experiment collapses to a single curve, given by $L_s/L_s^o=(1-\dot{\gamma}/\dot{\gamma}_c)^{-1/2}$. The reproduced results can be seen in,  Fig.~\ref{fig:ThndTr}a. However, the non-linear relationship of slip with shear rate and the universality of their boundary condition were only demonstrated for a single phase flow and as shown in Thalakkottor \& Mohseni \cite{Mohseni:13e} are valid for steady flow only. \textcolor{black}{In order to verify if their boundary condition is applicable to more complex flows, we perform molecular dyamics simulations of a single-phase corner flow and two-phase Couette flow.}
Schematic of the problem geometry is shown in Fig.~\ref{fig:Schematic} and the details of the problem setup and the MD simulations are given in the following section. The various fluid-wall properties used in this letter are tabulated in Table~\ref{tab:wll_fl_properties}. It is seen that Thompson \& Troian's scaling does not result in collapsed data for the moving contact line problem (Fig.~\ref{fig:ThndTr}b). In addition, it is observed that the slip length starts to diverge even though the local shear rate has not approached the critical value. Similar results can be shown in the case of a corner flow. This suggests that Thompson \& Troian's model while valid for single phase flows on a regular surface, has certain limitations when applied to problems such as a moving contact line and a corner flow.

\section{Experimental setup}

\begin{figure}[h!]
\centering
\begin{minipage}{0.45\linewidth}\begin{center}
 \includegraphics[width=0.9\linewidth]{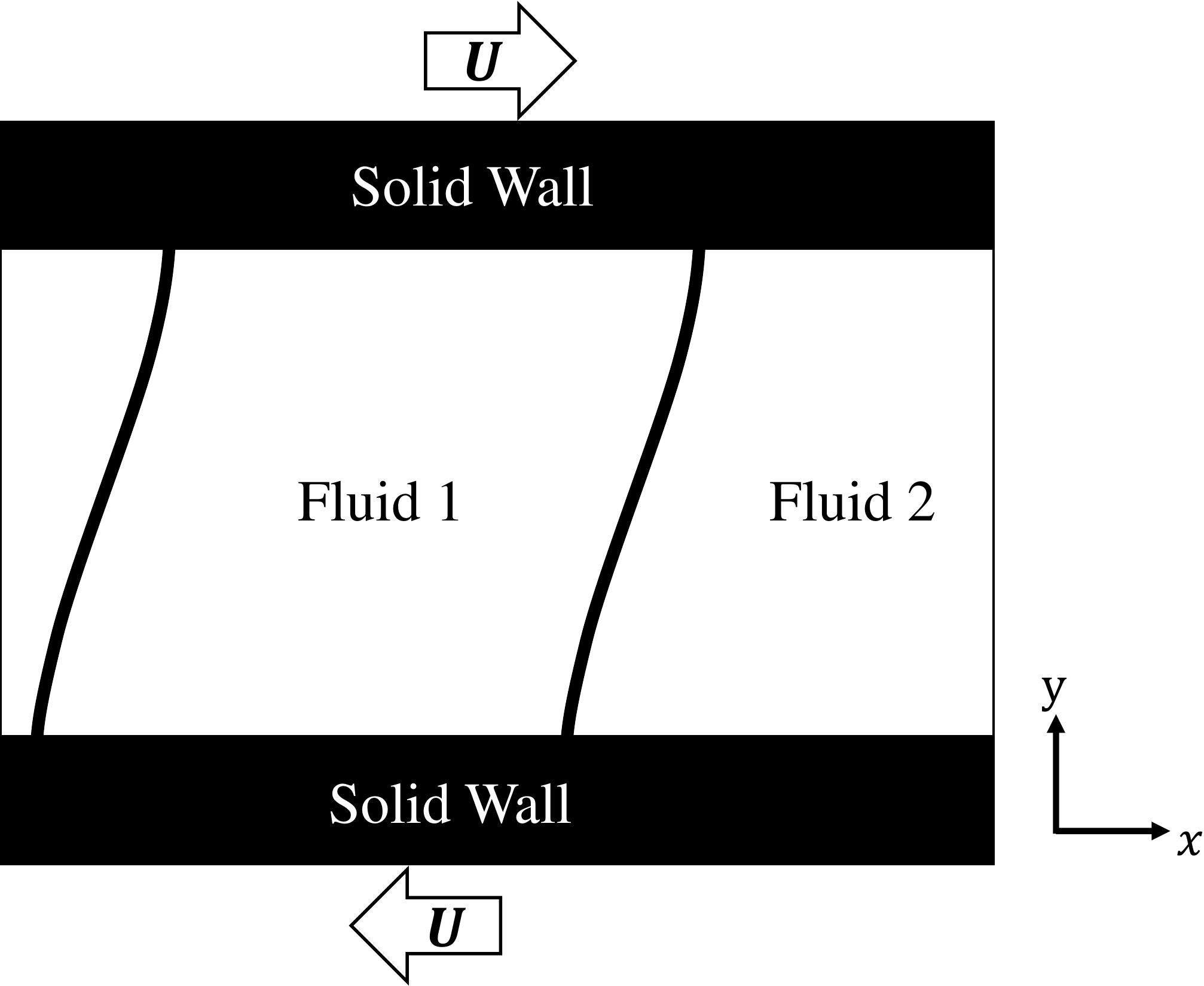}
\end{center}
\end{minipage}
\begin{minipage}{0.45\linewidth}\begin{center}
 \includegraphics[width=0.9\linewidth]{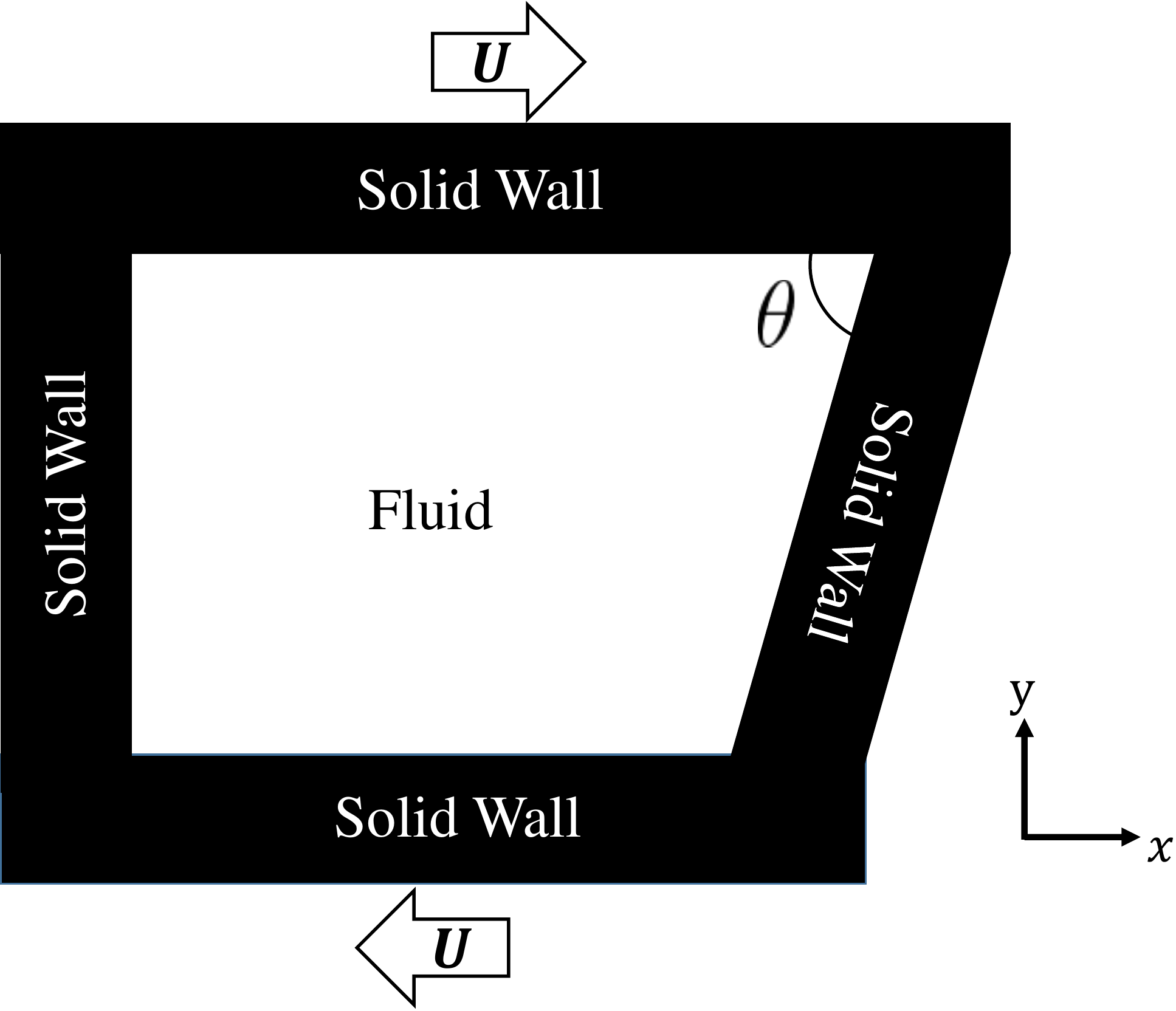}
\end{center}
\end{minipage} \\
\begin{minipage}{0.49\linewidth}\begin{center} (a) Moving contact line \end{center}\end{minipage}
\begin{minipage}{0.49\linewidth}\begin{center} (b) Corner flow\end{center}\end{minipage} 
\caption{{\bf Schematics of the problem geometry.} (a) A two-phase, two dimensional Couette flow problem, where the fluid channel measures $153\sigma\times144\sigma\times27.4\sigma$. The walls move in opposite directions with a speed $U=0.1\sigma/\tau$ and periodic boundary conditions are imposed along the $x$ and $z$ directions. (b) A two dimensional corner flow problem, where the corner flow is simulated by modelling a cavity flow with an inclined wall. The cavity measures $91\sigma\times72\sigma\times24.4\sigma$. The top and bottom walls move with a speed $U=0.1\sigma/\tau$, while the side walls are stationary. A periodic boundary condition is imposed along the $z$ direction, which is the out of the plane axis.}
\label{fig:Schematic}
\end{figure}

{\bf Problem description.} {\it Moving contact line problem.} In this letter we investigate slip at the triple contact point by simulating a two-phase Couette flow. Two immiscible fluids occupying equal volume are placed in between two parallel walls,  which move in opposite directions with a speed $U$ and are separated by a distance $H$. Periodic boundary conditions are imposed along the $x$ and $z$ directions. The schematic of the problem geometry is shown in Fig.~\ref{fig:Schematic}a. \\
{\it Corner flow problem.} We study slip at the corner point by simulating a cavity with an inclined side wall. Here, the top and bottom walls move in opposite directions with a speed $U$, while the side walls are stationary. Periodic boundary conditions are imposed along the $z$ direction. The schematic of the problem geometry is shown in Fig.~\ref{fig:Schematic}b. %In this study we focus on one of the corners.
\\

\noindent{\bf Molecular dynamics simulations.}
The numerical simulations are performed using molecular dynamics \cite{PlimptonS:95a}, where the pairwise interaction of molecules, separated by a distance $r$, is modelled by the Lennard Jones (LJ) potential
\begin{equation}
V^{LJ}=4\epsilon\left[\left(\frac{\sigma}{r}\right)^{12}-\left(\frac{\sigma}{r}\right)^{6}\right].
\end{equation}
Here, $\epsilon$ and $\sigma$  are the characteristic energy and length scales, respectively. The potential is zero for $r>r_c=2.5\sigma$, where $r_c$ is the cutoff radius. 
%
%Here fluid parameters corresponding to liquid Argon are chosen.

Each wall is comprised of at least two layers of molecules oriented along the $(111)$ plane of a face centered cubic (fcc) lattice, with the molecules fixed to their lattice site. \textcolor{black}{The fluid molecules are initialized on a} fcc lattice  whose spacing is chosen to obtain the desired density, with initial velocities randomly assigned subject to a fixed temperature.
The equilibrium state, has a temperature $T=1.1k_B/\epsilon$ and number density $\rho\approx0.81\sigma^{-3}$ for the corner flow problem, and $\rho\approx 0.73\sigma^{-3}$ for the moving contact line problem. The temperature is maintained  using a Langevin thermostat with a damping coefficient of $\Gamma=0.1\tau^{-1}$. The characteristic MD time unit is $\tau=\sqrt{m\sigma^{2}/\epsilon}$, where $m$ is the mass of the fluid molecule. The damping term is only applied to the $z$ direction to avoid biasing the flow.
In the case of the moving contact line problem the immiscibility of the two fluids is modelled by choosing appropriate LJ interaction parameters, such that the interatomic forces between them is predominantly repulsive. \textcolor{black}{For the results presented here these parameters are $\epsilon^{f1f2}=0.2\epsilon$, $\sigma^{f1f2}=3.0\sigma$ and $r_c=2.5\sigma$, which ensure a purely repulsive force. For simplicity the two fluids are assigned identical fluid properties.} \textcolor{black}{The model has been verified for additional cases of both corner flow and moving contact line problems, but they are not presented here for legibility reasons.}

% \noindent{\bf Production run and extracting data.}
 The equations of motion are numerically integrated using the Verlet algorithm with a time step $\Delta t=0.002\tau$.
The simulation is initially run until the flow equilibrates, after which spatial averaging is performed by dividing the fluid domain into square bins of size $\sim0.5\times1.0 \sigma$ along the $x$--$y$ plane, and extending through the entire width of the channel. In addition to spatial averaging, time averaging is done for a duration of $8000\tau$ for the moving contact line problem %\textcolor{blue}{check i probably changed it to time averaging}
. In the case of a non-wetting wall, averaging was done for an extended time of $16000\tau$ in order to  better resolve the data. For the corner flow problem time averaging is performed for a duration of $200000\tau$. 
%
%{\bf extracting data}
The velocity results presented in this paper are normalized by the wall velocity, $|U|=0.1\sigma\tau^{-1}$. In order to compute various quantities at the wall, a reference plane is defined at a distance of $\sigma^{wf}$ away from the wall lattice site. Data points at $2\sigma-3\sigma$ away from the contact point and corner point are excluded as the data is unresolved because of local fluid velocity approaching zero. %(or is it from the lattice location of the wall) %for which $Kn<1$
\\

%\\We attempt to verify, using molecular dynamics (MD) simulations, if their model is applicable to complex flows such as a single phase corner flow and two-phase Couette flow.}\\
%%%%%%%%%%%%%%%%%%%%%%%%%%%%%%%%%%%%%%%%%%%%%%%%%%%%%%%%%%%%%%%%%%%%

 %=======================================================================================================================================
 %%%%%%%%%%%%%%%%%%%%%%%%%%%%%%%%%%%%%%%%%%%%%%%%%%%%%%%%%%%%%%%%%%%%
%%  TABLES
 %%%%%%%%%%%%%%%%%%%%%%%%%%%%%%%%%%%%%%%%%%%%%%%%%%%%%%%%%%%%%%%%%%%%

 \begin{table}[h!]
 \begin{center}
\begin{tabular}{l*{6}{c}r}
  \hline
  \hline
  & Case & ~$\epsilon^{wf}/\epsilon$~ & ~$\sigma^{wf}/\sigma$~ & ~$\rho^w/\rho$~ \\
  \hline
  &1~ &~$1.0$~         & ~$1.0$~ & ~$1$~ \\
  &2~ &~$0.6$~         & ~$1.0$~ & ~$1$~ \\
  &3~ &~$0.6$~         & ~$0.75$~ & ~$4$~ \\
  &4~ &~$0.4$~         & ~$0.75$~ & ~$4$~ \\
\hline
\hline
\end{tabular}
\caption{\label{tab:wll_fl_properties} {\bf \textcolor{black}{Parameters for different cases of fluid-wall interface with slip increasing from Case 1 to 4.}} ~$\epsilon^{wf}$ and $\sigma^{wf}$ are the Lennard-Jones (LJ) parameters for fluid-wall interaction, $\rho^w/\rho$ is relative density of wall. $\epsilon^{wf}$ determines the extent of affinity of the wall molecules to the fluid molecules and is inversely related to slip length. $\sigma^{wf}$ corresponds to the molecular diameter or length scale associated with the LJ potential. An increase in its value leads to greater slip and vice-versa. Higher relative wall density means a smoother perceived surface leading to greater slip.} %Case 1 corresponds to  hydrophilic case}
\end{center}
\end{table}

%%%%%%%%%%%%%%%%%%%%%%%%%%%%%%%%%%%%%%%%%%%%%%%%%%%%%%%%%%%%%%%%%%%%%%%%%%%%%%%%%%%%%%%%%%%%%%%%%%%%%%%%%%%%%%%%
%%%%%%%%%%%%%%%%%%%%%%%%%%%%%%%%%%%%%%%		Setup	%%%%%%%%%%%%%%%%%%%%%%%%%%%%%%%%%%%%%%%%%%%%%%%%
%%%%%%%%%%%%%%%%%%%%%%%%%%%%%%%%%%%%%%%%%%%%%%%%%%%%%%%%%%%%%%%%%%%%%%%%%%%%%%%%%%%%%%%%%%%%%%%%%%%%%%%%%%%%%%%%

%%%%%%%%%%%%%%%%%%%%%%%%%%%%%%%%%%%%%%%%%%%%%%%%%%%%%%%%%%%%%%%%%%%%%%%%%%%%%%%%%%%%%%%%%%%%%%%%%%%%%%%%%%%%%%%%
%%%%%%%%%%%%%%%%%%%%%%%%%%%%%%%%%%%%%%%		T&T validation	%%%%%%%%%%%%%%%%%%%%%%%%%%%%%%%%%%%%%%%%%%%%%%%%
%%%%%%%%%%%%%%%%%%%%%%%%%%%%%%%%%%%%%%%%%%%%%%%%%%%%%%%%%%%%%%%%%%%%%%%%%%%%%%%%%%%%%%%%%%%%%%%%%%%%%%%%%%%%%%%%

%%%%%%%%%%%%%%%%%%%%%%%%%%%%%%%%%%%%%%%%%%%%%%%%%%%%%%%%%%%%%%%%%%%%%%%%%%%%%%%%%%%%%%%%%%%%%%%%%%%%%%%%%%%%%%%%
%%%%%%%%%%%%%%%%%%%%%%%%%%%%%%%%%%%%%%%		Slip model	%%%%%%%%%%%%%%%%%%%%%%%%%%%%%%%%%%%%%%%%%%%%%%%%
%%%%%%%%%%%%%%%%%%%%%%%%%%%%%%%%%%%%%%%%%%%%%%%%%%%%%%%%%%%%%%%%%%%%%%%%%%%%%%%%%%%%%%%%%%%%%%%%%%%%%%%%%%%%%%%%
\section{Generalized Slip model}
% \noindent{\bf General slip model derivation.}
The Maxwell slip model is revisited with an aim to capture the functionality associated with slip in a general steady flow. One of the reasons we resort to it is because of a lack of widely accepted kinetic models that are valid for liquids. Even though Maxwell's model  \cite{MaxwellJC:90a,LoebLB:34a,GombosiTI:94a} was established for rarefied gases, \textcolor{black}{it is illustrated here that an analogous formulation shows promise to identify model functionalities for slip modelling in liquids, or fluids in general.}  This is emphasized by the fact that the Navier slip model gives the same functionality of slip velocity with shear rate as Maxwell.
 Similar steps were taken by Thalakkottor \& Mohseni \cite{Mohseni:13e} to extend Maxwell's slip model to unsteady flows with success.
 
%  The mechanism of slip at the wall is expected to be similar for both liquids and gases.
\begin{figure}[h!]
\begin{center}
\includegraphics[width=0.30\textwidth]{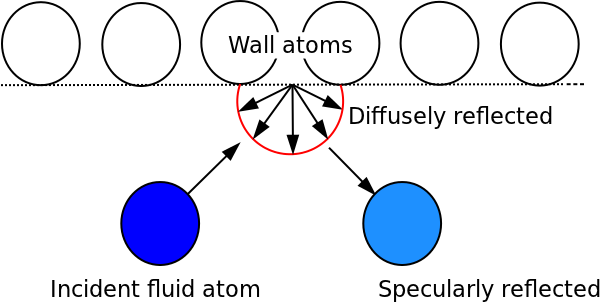}
 \caption{{\bf Schematic of diffusively and specularly reflected fluid molecules.}}
 \label{fig:Reflection}
 \end{center}
\end{figure}

For the derivation of the general slip model, a wall inclined at an arbitrary angle is considered. 
\textcolor{black}{Maxwell's theory states that the reflection of fluid molecules after interaction with the wall can be categorized into two types; namely, specular and diffusive reflection as shown in Fig.~\ref{fig:Reflection}. }
%
%
%
%
%%%%%%%%%%%%%%%%%%%%%%%%%%%%%%%%%%%%%%%%%%%%%%%%%%%%%%%%%%%%%%%%%%%%%%%%%%%%%%%%%%%%%%%%%%%%%%%%%%%%%%%%%%%%%%%%%%%%%%%%%%%%%%%%%%%%
\\

\noindent\textit{Diffusive reflection}~$\left(\boldsymbol{u}^{\text{diff}}\right)$:
The incident fluid molecules can be imagined as being adsorbed by the wall and then emitted into the fluid, such that the net velocity will be the same as that of the fluid being at rest with respect to the wall, that is
 \begin{equation}
{u}^{\text{diff}}(t_c^+)=\boldsymbol{s}\boldsymbol{\cdot}\boldsymbol{U}(t_c).
\label{eq:nd_1}
\end{equation}
Here, $t_c$ is the time of collision, $t_c^+$ is the \textcolor{black}{instantaneous time immediately} after collision with the wall, $\boldsymbol{U}$ is the wall velocity vector and ${\boldsymbol{s}}$ is the wall-tangent unit vector that is parallel to the flow direction. All the parameters are evaluated at the wall unless specified.\\

\noindent\textit{Specular reflection}~$\left(\boldsymbol{u}^{\text{spec}}\right)$:
\textcolor{black}{The incident fluid molecules undergo perfect elastic collision with the wall, such that there is no tangential momentum transfer with the wall. Therefore,}
\begin{align}
{u}^{\text{spec}} (t_c^+) &=\boldsymbol{s}\boldsymbol{\cdot} \boldsymbol{u}(t_c^-),
\label{eq:nd_2}
\end{align}
where $t_c^-$ is the instantaneous time immediately before collision. The velocity of the incident molecule is obtained by collision with a fluid molecule located at a distance away from the wall, such that the average distance travelled by it in a mean free time is $2/3\lambda$ in the $x$, $y$ and $z$ directions, where $\lambda$ is the mean free path. 
Hence, the incident velocity is calculated by performing a Taylor series expansion in space about the wall:
\begin{equation}
  {u}^{\text{spec}} (t_c^+) = 
  \boldsymbol{s\cdot}\boldsymbol{u}_{\frac{2}{3}\lambda}(t_c)=
  \boldsymbol{s\cdot}\boldsymbol{u}\left(t_c\right)- \frac{2}{3}\lambda\boldsymbol{s\cdot}\boldsymbol{\nabla} \boldsymbol{u}\left(t_c\right)\boldsymbol{\cdot}\boldsymbol{\Delta}.
\end{equation}
Here, $\boldsymbol{u}_{\frac{2}{3}\lambda}$ is the fluid velocity at a distance $2/3\lambda$ away from the wall,  $\boldsymbol{\Delta}$ is the direction vector of the incident fluid molecule. 

\textcolor{black}{Now, the momentum of diffusively reflected molecules together with specularly reflected molecules give the net momentum of the reflected molecules. The fractions of diffusive and specular molecules making up the net reflected molecules are determined by the tangential momentum accommodation coefficient ($\tilde{\sigma}$)}. 

Therefore, the net reflected velocity is written as
\begin{equation}
  {u}(t_c^+)=\tilde{\sigma} {u}^{\text{diff}}(t_c^+) + \left(1-\tilde{\sigma}\right){u}^{\text{spec}}(t_c^+).
\end{equation}

As the incident and reflected velocities together constitute the actual fluid molecules close to the surface, the average fluid velocity at the wall is given as the mean of the  velocity before and after collision
\begin{equation}
  {u}(t_c)=\frac{{u}(t_c^+) + {u}(t_c^-)}{2}.
\label{eq:nd_8}
\end{equation}
By substituting and simplifying, the slip velocity can be written as
\begin{equation}
 U_s = \frac{2}{3}\frac{(2-\tilde{\sigma})}{\tilde{\sigma}}\lambda\boldsymbol{{s}\cdot\nabla\boldsymbol{u}\cdot\boldsymbol{\Delta}},
\label{eq:nd_13}
\end{equation}
where ${U}_s=\boldsymbol{s\cdot}\boldsymbol{U}-\boldsymbol{s\cdot}\boldsymbol{u}$. The velocity gradient tensor can be written as the sum of a symmetric (strain rate) and anti-symmetric (rotation rate) tensor. We have verified that slip velocity is independent of the rotation rate tensor. This is because the rotation rate tensor does not cause any strain in the fluid element. Thus the above equation becomes
\begin{equation}
 U_s = L_s\boldsymbol{{s}\cdot(\nabla\boldsymbol{u}+\nabla\boldsymbol{u}^T)\cdot\boldsymbol{\Delta}}.
\label{eq:nd_14}
\end{equation}
Here, the coefficient $\frac{2}{3}\frac{\left(2-\tilde{\sigma}\right)}{\tilde{\sigma}}\lambda$, which is a measure of slip at the interface, is replaced by slip length, $L_s$. This allows the model to be applicable for liquids as well, for which the mean free path is not well defined.

If we consider a 2D case, then the tangential slip model simplifies to,
\begin{align}
 U_s&=L_s \left[\frac{\partial u}{\partial s} + \frac{\partial u}{\partial n}\right]. 
\label{eq:nd_18}
\end{align}
Here, $\boldsymbol{n}$ is the wall-normal unit vector.
 Comparing to the Navier/Maxwell slip model, it is seen that the slip velocity has an additional dependence on the linear strain rate. The primary difference between the Navier/Maxwell model and the general slip model presented here arises from the fact that in the case of the Navier/Maxwell slip models, the fluid velocity is only considered to change in the wall normal direction. Hence, all the terms in the velocity gradient tensor, except the one corresponding to the shear rate, reduce to zero. It can be concluded that for an arbitrary 3D flow the change in fluid velocity must not be limited to changes only in the wall normal direction and that the slip velocity is a function of the total strain rate along the wall. Thus, their model can be viewed as a limiting case of the more general velocity slip model presented here. However, in order to use the slip boundary condition, slip length needs to be known a priori. \textcolor{black}{This is addressed next.}

%%%%%%%%%%%%%%%%%%%%%%%%%%%%%%%%%%%%%%%%%%%%%%%%%%%%%%%%%%%%%%%%%%%%%%%%%%%%%%%%%%%%%%%%%%%%%%%%%%%%%%%%%%%%%%%%
%%%%%%%%%%%%%%%%%%%%%%%%%%%%%%%%%%%%%%%		Universal Curve		%%%%%%%%%%%%%%%%%%%%%%%%%%%%%%%%%%%%%%%%%%%%%%%%
%%%%%%%%%%%%%%%%%%%%%%%%%%%%%%%%%%%%%%%%%%%%%%%%%%%%%%%%%%%%%%%%%%%%%%%%%%%%%%%%%%%%%%%%%%%%%%%%%%%%%%%%%%%%%%%%
\section{Universal relationship for slip length}
 %%%%%%%%%%%%%%%%%%%%%%%%%%%%%%%%%%%%%%%%%%%%%%%%%%%%%%%%%%%%%%%%%%%%
 \begin{figure}[h!]
\centering
\begin{minipage}{0.45\linewidth}\begin{center}
 \includegraphics[width=0.9\linewidth]{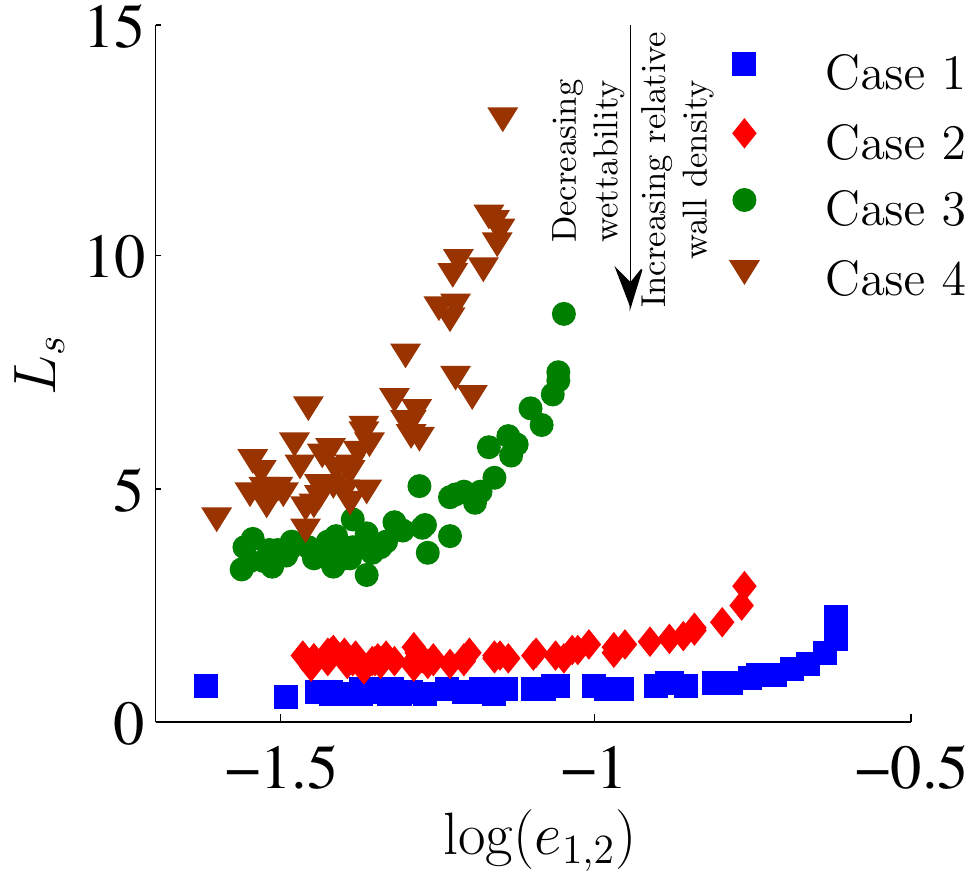}
\end{center}\end{minipage}
\begin{minipage}{0.45\linewidth}\begin{center}
 \includegraphics[width=0.9\linewidth]{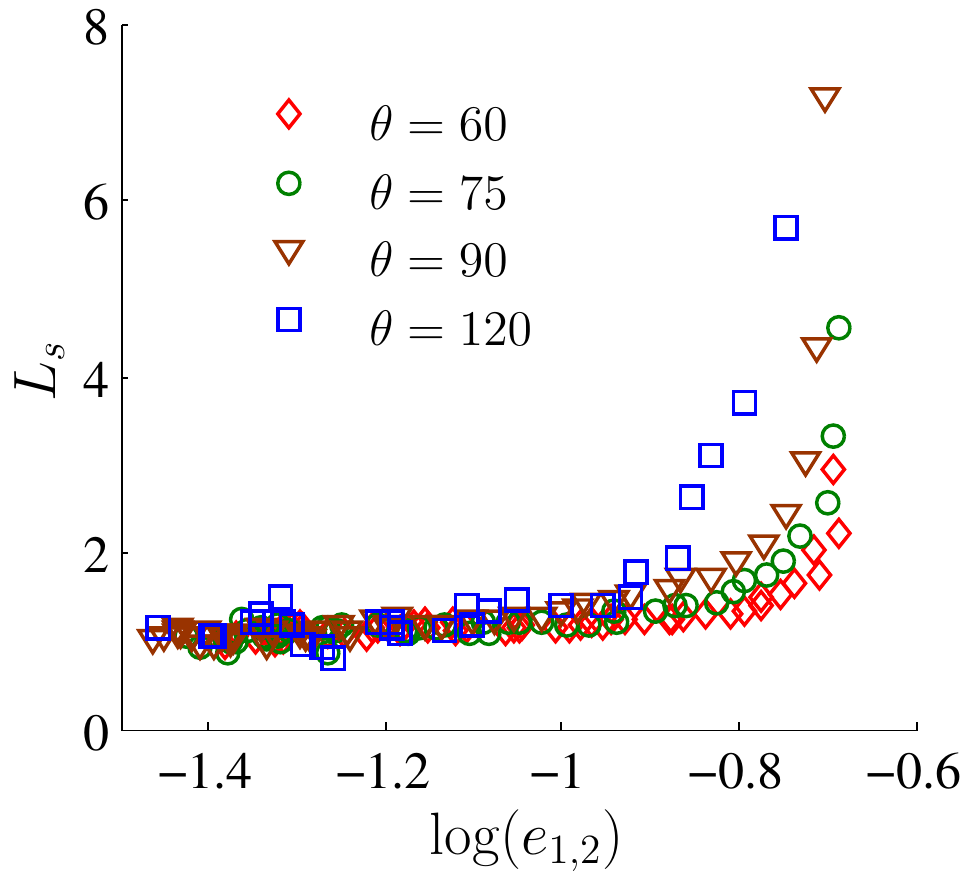}
\end{center}\end{minipage} \\
\begin{minipage}{0.49\linewidth}\begin{center} (a) moving contact line \end{center}\end{minipage} 
\begin{minipage}{0.49\linewidth}\begin{center} (b) corner flow \end{center}\end{minipage} 

\caption{{\bf Slip length versus principal strain rate, for different cases of interfacial properties and corner angles.} (a) Results for the trailing edge of a moving contact line problem are presented for four different cases of fluid-wall properties. (b) Results for a corner flow problem with wall moving towards the the corner (analogous to trailing edge) are presented for fluid-wall property corresponding to Case 2. Depending on whether we are considering the leading or trailing edge, the principal strain rate $e_1$ or $e_2$ are used, respectively. $e_{1,2}=(e_{xx}+e_{yy})/2\pm\sqrt{((e_{xx}-e_{yy})/2)^2+e_{xy}^2)}$, where $e_{xx}=\partial u/\partial x$, $e_{yy}=\partial v/\partial y$ and $e_{xy}=1/2(\partial u/\partial y+\partial v/\partial x)$. Results for additional cases are provided in Fig.~\ref{fig:leading_edge}.} 
\label{fig:Mod_slip}
\end{figure}
 
 %%%%%%%%%%%%%%%%%%%%%%%%%%%%%%%%%%%%%%%%%%%%%%%%%%%%%%%%%%%%%%%%%%%%

 \begin{figure}[h!]
\centering
\begin{minipage}{0.45\linewidth}\begin{center}
 \includegraphics[width=0.9\linewidth]{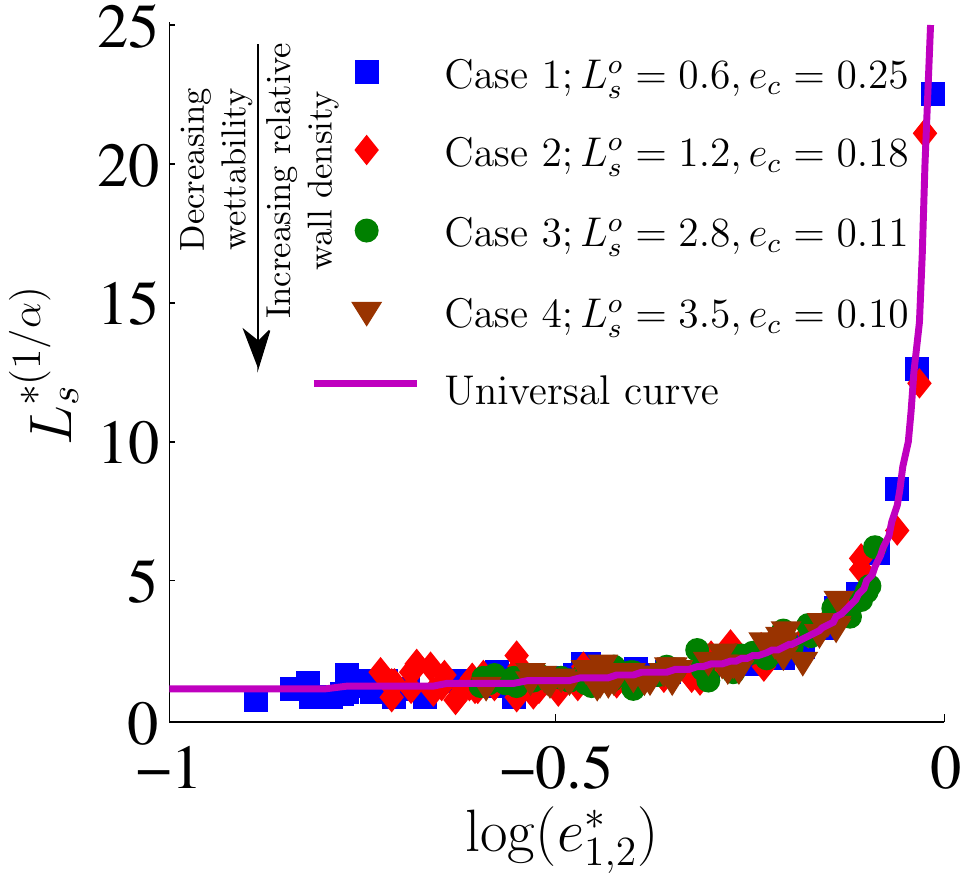}
\end{center}\end{minipage}
\begin{minipage}{0.45\linewidth}\begin{center}
 \includegraphics[width=0.9\linewidth]{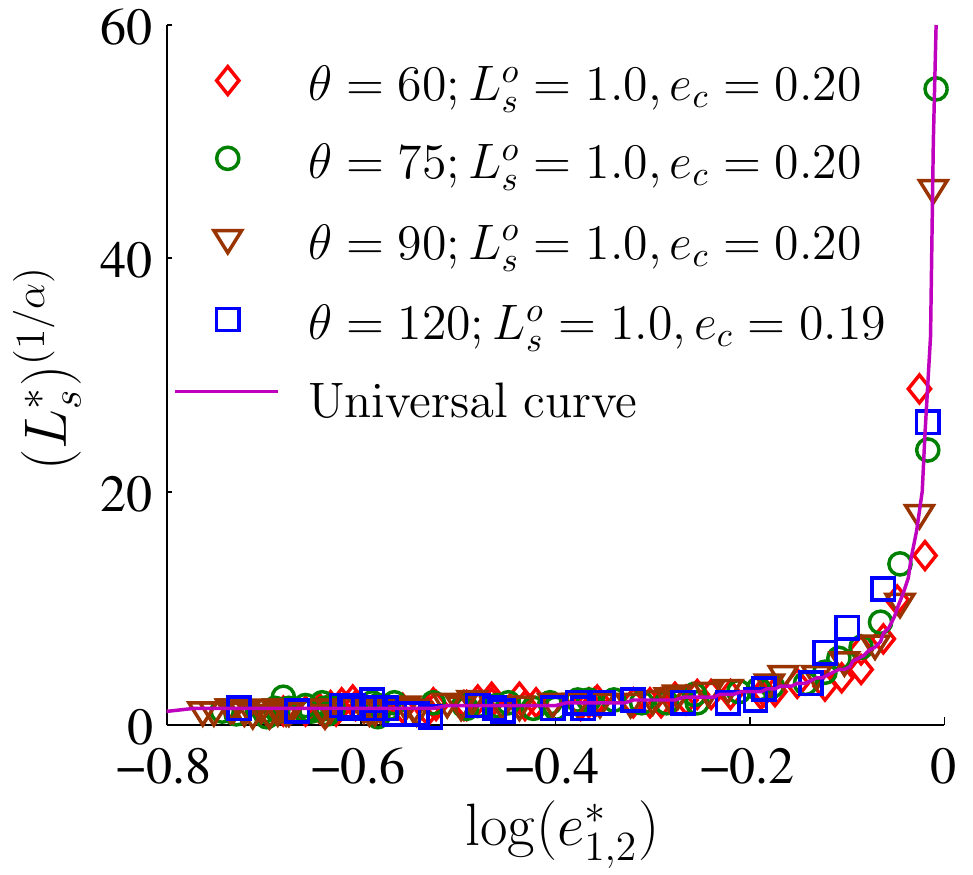}
\end{center}\end{minipage} \\
\begin{minipage}{0.49\linewidth}\begin{center} (a) moving contact line \end{center}\end{minipage} 
\begin{minipage}{0.49\linewidth}\begin{center} (b) corner flow \end{center}\end{minipage} 

\caption{{\bf Universal curve describing the slip along the wall.} Results are presented for, (a) the trailing edge of a moving contact line problem and (b) a corner flow problem with wall moving towards the corner. The universal curve is defined by $L_s^*=(1-e_{1,2}^*)^{\alpha}$, where $L_s^*=L_s/L_s^o$ and $e_{1,2}^*=e_{1,2}/e_{c}$. Additional results are provided in Fig.~\ref{fig:leading_edge_collap}.}
\label{fig:Mod_slip_collap}
\end{figure} 

%%%%%%%%%%%%%%%%%%%%%%%%%%%%%%%%%%%%%%%%%%%%%%%%%%%%%%%%%%%%%%%%%%%%
 
 \begin{figure}[h!]
\centering
\begin{minipage}{0.45\linewidth}\begin{center}
 \includegraphics[width=0.9\linewidth]{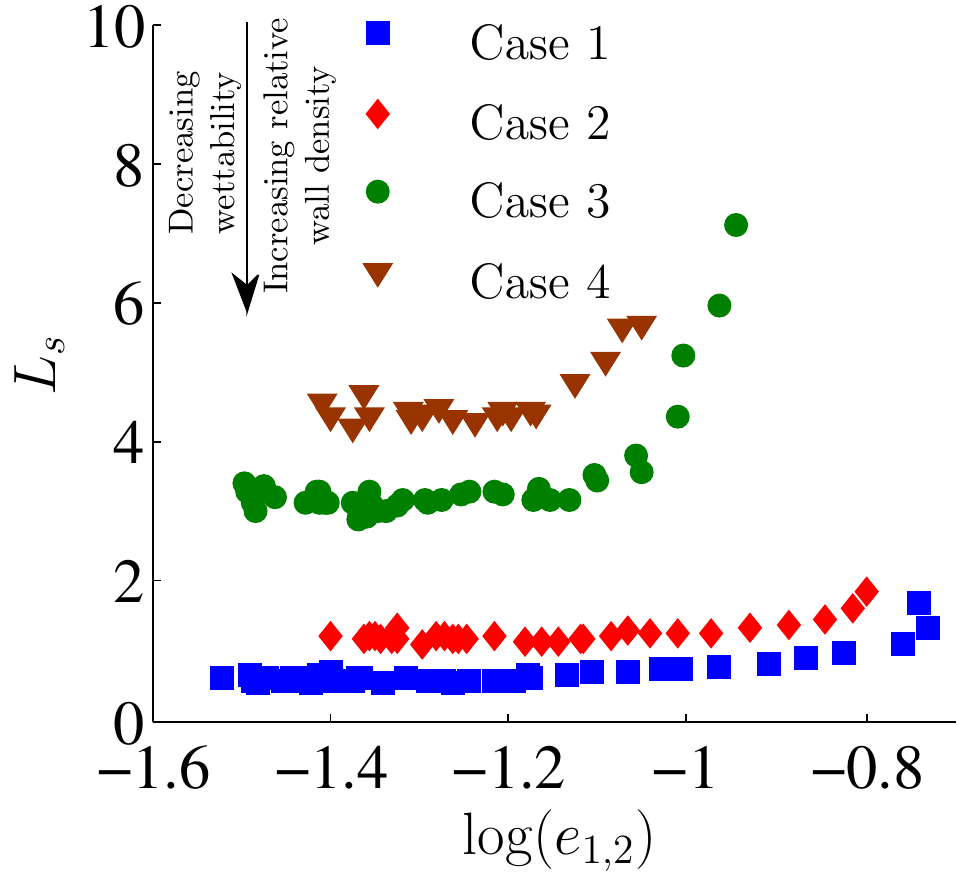}
\end{center}\end{minipage}
\begin{minipage}{0.45\linewidth}\begin{center}
 \includegraphics[width=0.9\linewidth]{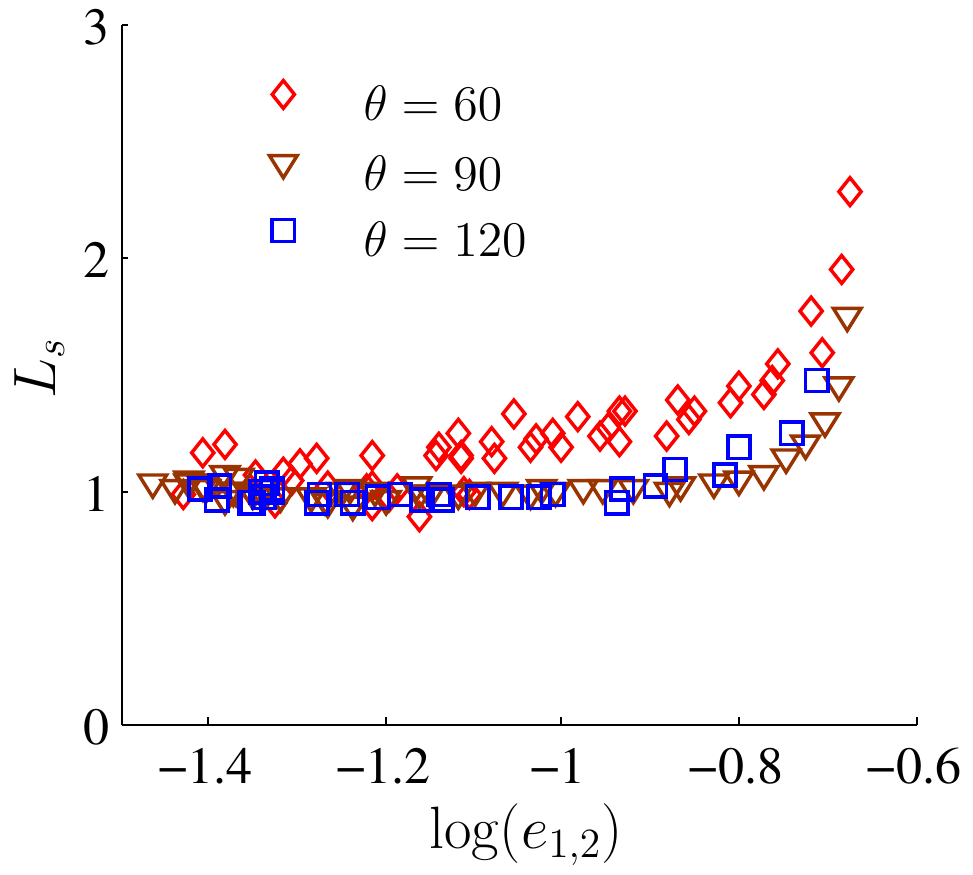}
\end{center}\end{minipage} \\
\begin{minipage}{0.49\linewidth}\begin{center} (a) moving contact line \end{center}\end{minipage}
\begin{minipage}{0.49\linewidth}\begin{center} (b) corner flow \end{center}\end{minipage} 

\caption{{\bf Slip length versus principal strain rate for additional cases.} (a) Results for the leading edge of a moving contact line problem are presented for four different cases of fluid-wall properties. (b) Results for a corner flow problem with wall moving away from the corner (analogous to leading edge) are presented for different corner angles and the fluid-wall  property corresponding to Case 2.}
\label{fig:leading_edge}
\end{figure}
 
%  %%%%%%%%%%%%%%%%%%%%%%%%%%%%%%%%%%%%%%%%%%%%%%%%%%%%%%%%%%%%%%%%%%%%
  
 \begin{figure}[h!]
\centering
\begin{minipage}{0.45\linewidth}\begin{center}
 \includegraphics[width=0.9\linewidth]{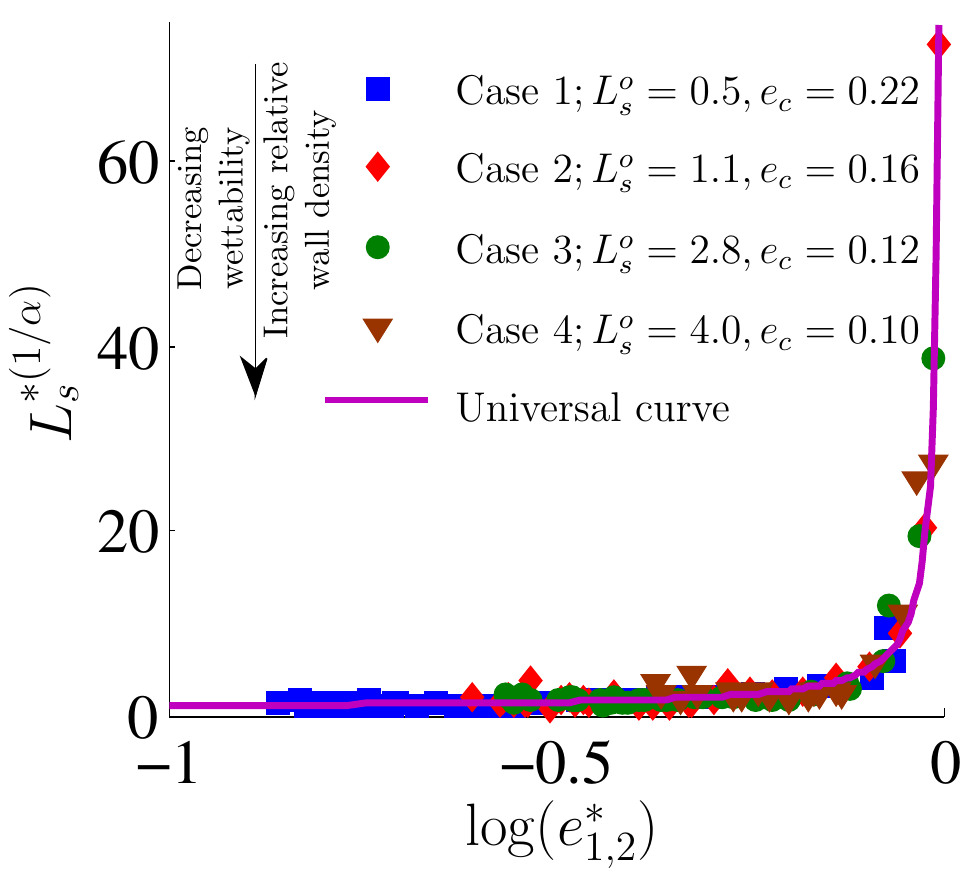}
\end{center}\end{minipage}
\begin{minipage}{0.45\linewidth}\begin{center}
 \includegraphics[width=0.9\linewidth]{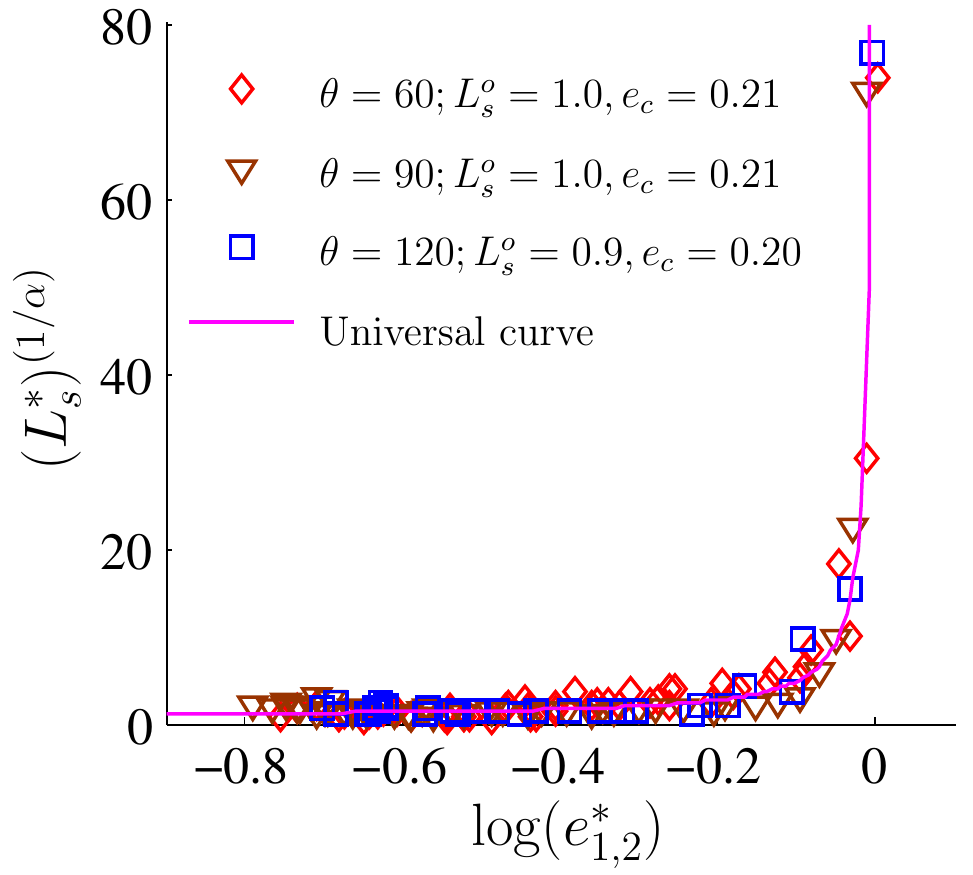}
\end{center}\end{minipage} \\
\begin{minipage}{0.49\linewidth}\begin{center} (a) moving contact line \end{center}\end{minipage}
\begin{minipage}{0.49\linewidth}\begin{center} (b) corner flow \end{center}\end{minipage} 

\caption{{\bf Universal curve describing slip along the wall for additional cases.} Results are presented for, (a) the leading edge of a moving contact line problem and (b) a corner flow problem with wall moving away from the corner.}
\label{fig:leading_edge_collap}
\end{figure}

\begin{figure}[h!]
\centering
 \includegraphics[width=0.45\linewidth]{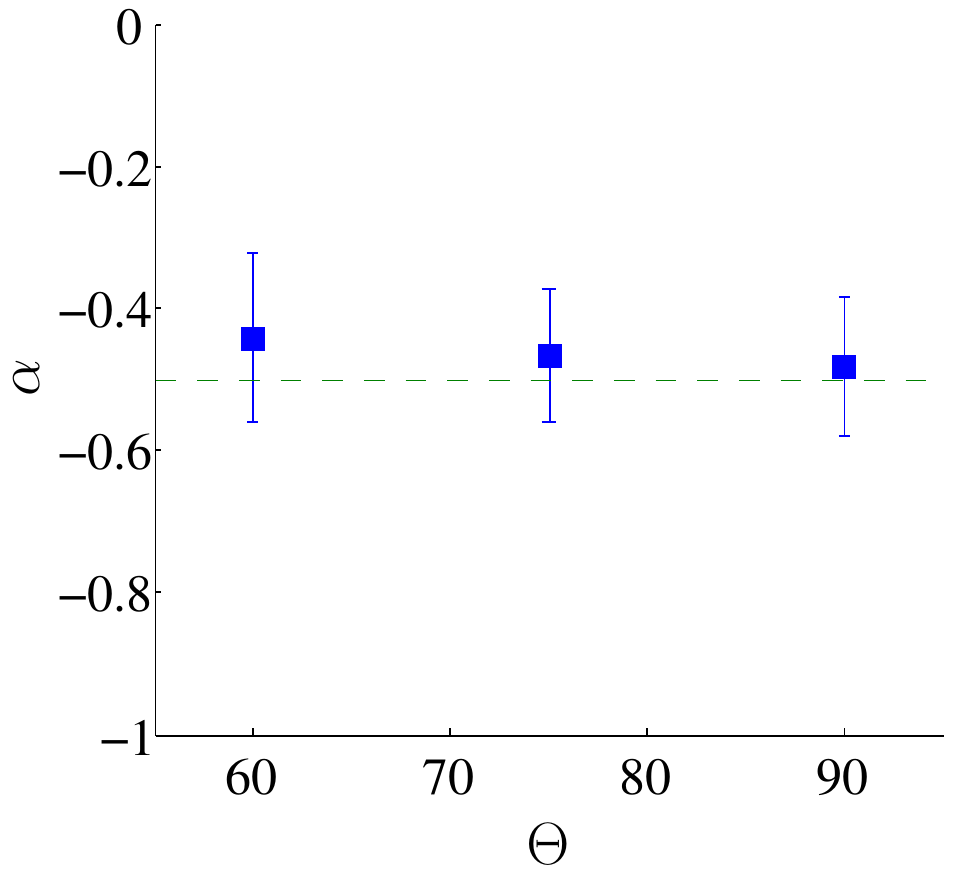}
\caption{\textcolor{black}{{\bf $\alpha$ for different contact angles $\theta$.} The parameter $\alpha$ is seen to be approximately $-0.5$, when the results are computed at a reference plane $0.5\sigma^{wf}$ away from the wall lattice site. This is consistent with Thompson \& Troian. If the values are not evaluated at the reference plane then $\alpha$ could show corner/contact angle dependency.}}

\label{fig:alphadist}
\end{figure}

The above findings suggest that for an arbitrary flow, slip length must not just be a function of shear rate \cite{TroianSM:97a}, but rather a function of a flow parameter that captures the total strain rate experienced in a fluid element. One such parameter is the principal strain rate, which represents the maximum and minimum strain rate in a fluid element. In Fig.~\ref{fig:Mod_slip}, the variation of slip length versus the local principal strain rate near the trailing edge of a moving contact line and a corner point is plotted. We observe a non-linear relationship between slip length and principal strain rate, the functional behaviour of which suggests the existence of a universal curve. Scaling slip length by its asymptotic value and the principal strain rate ($e_{1,2}$) by its critical value ($e_{c}$), the data collapses onto a single curve, described by $L_s=L_s^o(1-e_{1,2}/e_{c})^{\alpha}$, see in Fig.~\ref{fig:Mod_slip_collap}. This relation implies that close to the critical principal strain rate, slip lengths would approach macroscopic values and at the critical value one would observe perfect slip. 
 \textcolor{black}{This is analogous to material failure by plastic yielding or fracture where the limiting stress is a function of the principal stresses \cite{NadaiA:50a}.}
\textcolor{black}{The MD results show that the value of $\alpha\approx -0.5$ (Fig.~\ref{fig:alphadist}).} The validity of the universal relation is shown for additional cases in Fig.~\ref{fig:leading_edge} and Fig.~\ref{fig:leading_edge_collap}.
In the case of a steady incompressible single phase Couette flow the principal strain rate is equal to half the shear rate. 
\textcolor{black}{Hence, we can say that Thompson \& Troian's \cite{TroianSM:97a} slip model is the zero-linear-strain-rate limit of this universal relation and in turn, Navier's/Maxwell's \cite{NavierCLMH:1823a,MaxwellJC:90a} model is the low-shear-rate limit of Thomspon \& Troian's model.}

%%%%%%%%%%%%%%%%%%%%%%%%%%%%%%%%%%%%%%%%%%%%%%%%%%%%%%%%%%%%%%%%%%%%
 
\section{Conclusion}

To summarise, we presented a unified boundary condition that describes slip at the boundary for a wide range of steady flow problems. The boundary condition captures the physics associated with complex problems, such as single-phase corner flows and two-phase moving contact lines, while also being consistent with the slip models of Navier/Maxwell and Thompson \& Troian, for simpler flows. Our results suggest that the moving contact line and corner flow problems, both of which exhibit boundary singularities, are fundamentally similar in nature \textcolor{black}{and are governed by the flow conditions presented here.}

%=======================================================================================================================================
%=======================================================================================================================================
\section*{Acknowledgement}
%=======================================================================================================================================
%=======================================================================================================================================
This research was supported by the Office of Naval Research. We would also like to thank Dr. A. C. DeVoria for his helpful discussion and comments.

%=======================================================================================================================================

% \clearpage
% \newpage
%\printbibliography
% \bibliography{RefA1.bib}
%  \bibliography{RefA1}

\begin{thebibliography}{10}
\expandafter\ifx\csname url\endcsname\relax
  \def\url#1{\texttt{#1}}\fi
\expandafter\ifx\csname urlprefix\endcsname\relax\def\urlprefix{URL }\fi
\providecommand{\bibinfo}[2]{#2}
\providecommand{\eprint}[2][]{\url{#2}}

\bibitem{GoldsteinS:38a}
\bibinfo{author}{Goldstein, S.}
\newblock \emph{\bibinfo{title}{Modern Development in Fluid Dynamics}},
  vol.~\bibinfo{volume}{2}, \bibinfo{pages}{676--680}
  (\bibinfo{publisher}{Clarendon Press Oxford}, \bibinfo{year}{1938}).

\bibitem{GoldsteinS:69a}
\bibinfo{author}{Goldstein, S.}
\newblock \bibinfo{title}{Fluid mechanics in first half of this century}.
\newblock \emph{\bibinfo{journal}{Ann. Rev. Fluid Mech.}}
  \textbf{\bibinfo{volume}{1}}, \bibinfo{pages}{1--28} (\bibinfo{year}{1969}).

\bibitem{VinogradovaOI:99a}
\bibinfo{author}{Vinogradova, O.}
\newblock \bibinfo{title}{Slippage of water over hydrophobic surfaces}.
\newblock \emph{\bibinfo{journal}{Int. J. Miner. Process}}
  \textbf{\bibinfo{volume}{56}}, \bibinfo{pages}{31--60}
  (\bibinfo{year}{1999}).

\bibitem{NavierCLMH:1823a}
\bibinfo{author}{Navier, C.}
\newblock \bibinfo{title}{Memoire sur les lois du mouvement des fluides}.
\newblock \emph{\bibinfo{journal}{Memoires de l Academie Royale des Sciences de
  l Instituede France}} \textbf{\bibinfo{volume}{6}}, \bibinfo{pages}{389--440}
  (\bibinfo{year}{1823}).

\bibitem{MaxwellJC:90a}
\bibinfo{author}{Maxwell, J.}
\newblock \bibinfo{title}{The scientific papers of {J}ames {C}lerk {M}axwell}.
\newblock vol.~\bibinfo{volume}{V2}, \bibinfo{pages}{703--711}
  (\bibinfo{year}{1890}).

\bibitem{TroianSM:97a}
\bibinfo{author}{Thompson, P.} \& \bibinfo{author}{Troian, S.}
\newblock \bibinfo{title}{A general boundary condition for liquid flow at solid
  surfaces}.
\newblock \emph{\bibinfo{journal}{Nature}} \textbf{\bibinfo{volume}{389}},
  \bibinfo{pages}{360--362} (\bibinfo{year}{1997}).

\bibitem{QianT:06a}
\bibinfo{author}{Qian, T.}, \bibinfo{author}{Wang, X.-P.} \&
  \bibinfo{author}{Sheng, P.}
\newblock \bibinfo{title}{Molecular hydrodynamics of the moving contact line in
  two-phase immiscible flows}.
\newblock \emph{\bibinfo{journal}{Comm. in Computational Phys.}}
  \textbf{\bibinfo{volume}{1}}, \bibinfo{pages}{1--52} (\bibinfo{year}{2006}).

\bibitem{HuhC:71a}
\bibinfo{author}{Huh, C.} \& \bibinfo{author}{Scriven, L.}
\newblock \bibinfo{title}{Hydrodynamic model of steady movement of a
  solid/liquid/fluid contact line}.
\newblock \emph{\bibinfo{journal}{J. Colloid Interface Sci.}}
  \textbf{\bibinfo{volume}{35}}, \bibinfo{pages}{85} (\bibinfo{year}{1971}).

\bibitem{Hocking:82a}
\bibinfo{author}{Hocking, L.} \& \bibinfo{author}{Rivers, A.}
\newblock \bibinfo{title}{The spreading of a drop by capillary action}.
\newblock \emph{\bibinfo{journal}{J. Fluid. Mech.}}
  \textbf{\bibinfo{volume}{121}}, \bibinfo{pages}{425--442}
  (\bibinfo{year}{1982}).

\bibitem{Dussan:79a}
\bibinfo{author}{Dussan, E.}
\newblock \bibinfo{title}{On the spreading of liquids on solid surfaces:
  {S}tatic and dynamic contact lines}.
\newblock \emph{\bibinfo{journal}{\ARFM}} \textbf{\bibinfo{volume}{11}},
  \bibinfo{pages}{371--400} (\bibinfo{year}{1997}).

\bibitem{ThompsonP:89a}
\bibinfo{author}{Thompson, P.} \& \bibinfo{author}{Robbins, M.}
\newblock \bibinfo{title}{Simulations of contact line motion: slip and the
  dynamic contact angle}.
\newblock \emph{\bibinfo{journal}{Phys. Rev. Lett.}}
  \textbf{\bibinfo{volume}{63}}, \bibinfo{pages}{766--769}
  (\bibinfo{year}{1989}).

\bibitem{MoffattHK:64a}
\bibinfo{author}{Moffatt, H.}
\newblock \bibinfo{title}{Viscous and resistive eddies near a sharp corner}.
\newblock \emph{\bibinfo{journal}{Journal of Fluid Mechanics}}
  \textbf{\bibinfo{volume}{18}}, \bibinfo{pages}{1--18} (\bibinfo{year}{1964}).
\newblock
  \urlprefix\url{http://journals.cambridge.org/article_S0022112064000015}.

\bibitem{KoplikJ:95a}
\bibinfo{author}{Koplik, J.} \& \bibinfo{author}{Banavar, J.}
\newblock \bibinfo{title}{Corner flow in the sliding plate problem}.
\newblock \emph{\bibinfo{journal}{Physics of Fluids}}
  \textbf{\bibinfo{volume}{7}}, \bibinfo{pages}{3118--3125}
  (\bibinfo{year}{1995}).

\bibitem{RichardsonS:73a}
\bibinfo{author}{Richardson, S.}
\newblock \bibinfo{title}{On the no-slip boundary condition}.
\newblock \emph{\bibinfo{journal}{\JFM}} \textbf{\bibinfo{volume}{59}},
  \bibinfo{pages}{707--719} (\bibinfo{year}{1973}).

\bibitem{Mohseni:13e}
\bibinfo{author}{Thalakkottor, J.} \& \bibinfo{author}{Mohseni, K.}
\newblock \bibinfo{title}{Analysis of slip in a flow with an oscillating wall}.
\newblock \emph{\bibinfo{journal}{Phys. Rev. E}} \textbf{\bibinfo{volume}{87}}
  (\bibinfo{year}{2013}).

\bibitem{PlimptonS:95a}
\bibinfo{author}{Plimpton, S.}
\newblock \bibinfo{title}{Fast parallel algorithms for short-range molecular
  dynamics}.
\newblock \emph{\bibinfo{journal}{\JCP}} \textbf{\bibinfo{volume}{117}},
  \bibinfo{pages}{1--19} (\bibinfo{year}{1995}).

\bibitem{LoebLB:34a}
\bibinfo{author}{Loeb, L.}
\newblock \emph{\bibinfo{title}{The Kinetic Theory of Gases}}
  (\bibinfo{publisher}{McGraw-Hill Book Company}, \bibinfo{year}{1934}).

\bibitem{GombosiTI:94a}
\bibinfo{author}{Gombosi, T.}
\newblock \emph{\bibinfo{title}{Gaskinetic theory}}
  (\bibinfo{publisher}{Cambridge University Press}, \bibinfo{year}{1994}).

\bibitem{NadaiA:50a}
\bibinfo{author}{Nadai, A.}
\newblock \emph{\bibinfo{title}{Theory of Flow and Fracture of Solids}}
  (\bibinfo{publisher}{McGraw-Hill}, \bibinfo{address}{New york},
  \bibinfo{year}{1950}), \bibinfo{edition}{1} edn.

\end{thebibliography}
%  \bibliographystyle{naturemag}
 \newcommand{\AIAAJ}{AIAA J.} \newcommand{\AIAAP}{AIAA Paper}
  \newcommand{\ARMA}{Archive for Rational Mechanics and Analysis}
  \newcommand{\ASMEJFE}{J. Fluids Eng., Trans. ASME} \newcommand{\ASR}{Applied
  Scientific Research} \newcommand{\CF}{Computers Fluids}
  \newcommand{\CJFAS}{Can. J. Fish. Aquat. Sci.}
  \newcommand{\ETFS}{Experimental Thermal and Fluid Science}
  \newcommand{\EF}{Experiments in Fluids} \newcommand{\FDR}{Fluid Dynamics
  Research} \newcommand{\IJHMT}{Int. J. Heat Mass Transfer}
  \newcommand{\JASA}{J. Acoust. Soc. Am.} \newcommand{\JCP}{J. Comp. Physics}
  \newcommand{\JEB}{J. Exp. Biol.} \newcommand{\JFM}{J. Fluid Mech.}
  \newcommand{\JMP}{J. Math. Phys.} \newcommand{\JSC}{J. Scientific Computing}
  \newcommand{\JSP}{J. Stat. Phys.} \newcommand{\JSV}{J. of Sound and
  Vibration} \newcommand{\MC}{Mathematics of Computation}
  \newcommand{\MWR}{Monthly Weather Review} \newcommand{\PAS}{Prog. in
  Aerospace. Sci.} \newcommand{\PCPS}{Proc. Camb. Phil. Soc.}
  \newcommand{\PD}{Physica D} \newcommand{\PRA}{Physical Rev. A}
  \newcommand{\PRE}{Physical Rev. E} \newcommand{\PRL}{Phys. Rev. Lett.}
  \newcommand{\PF}{Phys. Fluids} \newcommand{\PFA}{Phys. Fluids A.}
  \newcommand{\PL}{Phys. Lett.} \newcommand{\PRSLA}{Proc. R. Soc. Lond. A}
  \newcommand{\SIAMJMA}{SIAM J. Math. Anal.} \newcommand{\SIAMJNA}{SIAM J.
  Numer. Anal.} \newcommand{\SIAMJSC}{SIAM J. Sci. Comput.}
  \newcommand{\SIAMJSSC}{SIAM J. Sci. Stat. Comput.}
  \newcommand{\TCFD}{Theoret. Comput. Fluid Dynamics} \newcommand{\ZAMM}{ZAMM}
  \newcommand{\ZAMP}{ZAMP} \newcommand{\ICASER}{ICASE Rep. No.}
  \newcommand{\NASACR}{NASA CR} \newcommand{\NASATM}{NASA TM}
  \newcommand{\NASATP}{NASA TP} \newcommand{\ARFM}{Ann. Rev. Fluid Mech.}
  \newcommand{\WWW}{from {\tt www}.} \newcommand{\CTR}{Center for Turbulence
  Research, Annual Research Briefs} \newcommand{\vonKarman}{von Karman
  Institute for Fluid Dynamics Lecture Series}

\end{document}